\documentclass[pra,print,superscriptaddress,twocolumn]{revtex4}
\usepackage[dvips]{graphics,color}
\usepackage{mathrsfs}
\usepackage{amsfonts}
\usepackage{amsmath}
\usepackage{amssymb}
\usepackage{graphicx}
\usepackage{multirow}
\usepackage{color}
\usepackage[colorlinks,citecolor=blue]{hyperref}
\usepackage{times}

\begin{document}

\title{Synthetic gauge field and chiral physics on two-leg superconducting circuits}

\author{Xin Guan}
\affiliation{State Key Laboratory of Quantum Optics and Quantum Optics Devices, Institute
of Laser spectroscopy, Shanxi University, Taiyuan 030006, China}

\author{Yanlin Feng}
\affiliation{Shandong Provincial Engineering and Technical Center of Light Manipulations
and Shandong Provincial Key Laboratory of Optics and Photonic Device, School
of Physics and Electronics, Shandong Normal University, Jinan 250014, China}

\author{Zheng-Yuan Xue}
\email{zyxue83@163.com}
\affiliation{Guangdong Provincial Key Laboratory of Quantum Engineering and Quantum Materials, GPETR Center for Quantum Precision Measurement, and School of Physics and Telecommunication Engineering, South China Normal University, Guangzhou 510006, China}
\affiliation{Frontier Research Institute for Physics, South China Normal University, Guangzhou 510006, China}

\author{Gang Chen}
\email{chengang971@163.com}
\affiliation{State Key Laboratory of Quantum Optics and Quantum Optics Devices, Institute
of Laser spectroscopy, Shanxi University, Taiyuan 030006, China}
\affiliation{Collaborative Innovation Center of Extreme Optics, Shanxi University,
Taiyuan, Shanxi 030006, China}
\affiliation{Collaborative Innovation Center of Light Manipulations and Applications,
Shandong Normal University, Jinan 250358, China}

\author{Suotang Jia}
\affiliation{State Key Laboratory of Quantum Optics and Quantum Optics Devices, Institute
of Laser spectroscopy, Shanxi University, Taiyuan 030006, China}
\affiliation{Collaborative Innovation Center of Extreme Optics, Shanxi University,
Taiyuan, Shanxi 030006, China}

\date{\today}

\begin{abstract}
Gauge field is essential for exploring novel phenomena in modern physics. However, it has not been realized in the recent breakthrough experiment about two-leg superconducting circuits with transmon qubits [Phys. Rev. Lett. 123, 050502 (2019)]. Here we present an experimentally-feasible method to achieve the synthetic gauge field by introducing ac microwave driving in each qubit. In particular, the effective magnetic flux per plaquette achieved can be tuned independently by properly choosing the driving phases. Moreover, the ground-state chiral currents for the single- and two-qubit excitations are obtained and the Meissner-vortex phase transition is found. In the Meissner phase, the ground-state chiral current increases as the magnetic flux increases, while it decreases in the vortex phase. In addition, the chiral dynamics that depends crucially on the initial state of the system is also revealed. Finally, the possible experimental observations of the chiral current and dynamics are addressed. Therefore, our results provide a new route to explore novel many-body properties induced by the interplay of gauge field, two-leg hoppings and interaction of photons on superconducting circuits.
\end{abstract}

\maketitle

\section{Introduction}

Due to their long coherence time, fine tunability and high-precision measurement \cite{XG17,ZL13}, superconducting circuits have emerged as a promising platform for processing quantum information \cite{GW17,AB20} and quantum computing \cite{FA19,IB11}, as well as implementing quantum simulation \cite{SS13, IMG14}. The recent quantum-simulation experiments have attracted great attention on fundamental many-body physics \cite{IC20}, such as magnets \cite{DSM15,AK17}, localizations \cite{AAH17,ML18}, molecular energies \cite{PJ16}, anyonic braiding statistics \cite{YPZ16,CS18}, topological magnon insulator \cite{WC19}, strongly-correlated quantum walks
\cite{ZY19}, and dissipatively-stabilized Mott insulator \cite{RM19} and quantum phase transition \cite{MF17}. Notice that the observed many-body physics are mainly based on a chain of superconducting circuits. In a recent breakthrough experiment, a two-leg superconducting circuits with $24$ transmon qubits has been reported and the  single- and double-excitation  dynamics has been observed \cite{YY19}. This experiment opens a new route to explore exotic many-body physics \cite{SN09,HHL10,FQP13,SU15,JPL16,SG18,RN18,ZZ19,RAS19,XL13}, which can be induced by the competition between interleg and intraleg hoppings and strong interaction of photons on superconducting circuits.

On the other hand, the gauge field is essential for a wide range of research from high energy physics \cite{GL19} and cosmology \cite{DB04} to ultracold atoms \cite{JD11,NG14,DWZ18} and condensed-matter physics \cite{EW16}. Meanwhile, on superconducting circuits, the synthetic gauge field was firstly proposed \cite{ypwang1,ypwang2} and realized \cite{PR17} in one unit cell by modulating the qubit couplings, and moreover, its induced chiral spin clusters have been achieved \cite{DWW19}. It is natural to ask an interesting question about how to achieve synthetic gauge fields in a two-leg superconducting qubit lattice with many unit cells. If realized, what interesting observable physics will occur?

In this paper, we present a feasible scheme to achieve the simulation of synthetic gauge fields on two-leg superconducting circuits. In contrast to the previous schemes \cite{ypwang1,ypwang2,PR17}, here we introduce an ac driving on each transmon qubit through the flux-bias line. More importantly, the realized synthetic magnetic flux per plaquette can be tuned independently by controlling the driving phases, which is better than the previous realizations in the other quantum simulation systems, such as ultracold atoms \cite{ATOM-MA,ATOM-MA1,ATOM-MA2, ATOM-BKS, ATOM-MM,ATOM-HM}, photonic \cite{PC1,PC2,PC3,PC4}, acoustics \cite{XW19}, ion trap \cite{IT}. Based on the realized synthetic magnetic flux, the ground-state chiral current with single- and two-excitations are obtained and the Meissner-vortex quantum phase transition is also found. In the Meissner phase, the ground-state chiral current increases as the magnetic flux increases, while it decreases in the vortex phase. The chiral dynamics that depends crucially on the initial state of the system is also revealed. Finally, the possible experimental observations of the chiral current and dynamics are also addressed. Therefore, our results provide a new way to explore rich many-body phenomena induced by the interplay of gauge field, two-leg hoppings and interaction of photons on superconducting circuits.

This paper is organized as follows. In Sec.~\ref{section2}, we realize the synthetic gauge field tuned independently. In Secs.~\ref{section3} and \ref{section4}, we discuss the ground-state chiral current and chiral dynamics with single- and two-excitations, respectively. In Sec.~\ref{section5}, we present the possible experimental observations. The conclusions are given in Sec.~\ref{section6}.\newline

\begin{figure}[t]
\includegraphics[width=8cm]{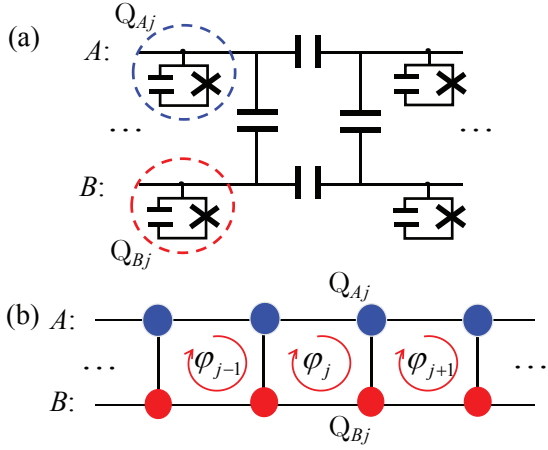}\newline
\caption{ Schematic diagram of two-leg (labeled respectively by $A$ and $B $) superconducting circuits with the transmon qubits. All transmon qubits are coupled with their nearest-neighbor sites by capacitors. (a) Illustration of the detail circuit of a unit plaquette. Q$_{\protect\nu j}$ denotes the qubit at the $j$th site on the $\protect\nu $th leg. (b)
Schematic diagram of the two-leg lattice with an effective magnetic flux per
plaquette which can be controlled independently. The blue and red solid
spheres indicate respectively the transmon qubits at the $A$ and $B$ legs.}
\label{Two}
\end{figure}
\section{Synthetic gauge field}\label{section2}

As shown in Fig.~\ref{Two}(a), we consider the same experimental setup about
two-leg superconducting circuits with the transmon qubits \cite{YY19}, whose
dynamics is governed by a Bose-Hubbard ladder Hamiltonian%
\begin{eqnarray}
\hat{H}_{BH}&=&\sum_{\nu j}\omega _{\nu j}^{0}\hat{a}_{\nu j}^{\dag }\hat{a}_{\nu
j} +\sum_{\nu j}\frac{V_{\nu j}}{2}\hat{n}_{\nu j}\left( \hat{n}_{\nu j}-1\right) \notag\\
&&+\sum_{\nu j}\left( g_{\nu j}\hat{a}_{\nu (j-1)}^{\dag }\hat{a}_{\nu j}+%
\text{H.c.}\right) \notag\\
&& +\sum_{j}\left( \tilde{g}_{j}\hat{a}_{Aj}^{\dag }\hat{a}_{Bj}+\text{H.c.}\right),  \label{H}
\end{eqnarray}
where $j$ is the number of the rung, $\nu \in \{A, B\}$ labels the leg,
the operator $\hat{a}_{\nu j}^{\dag }$ ($\hat{a}_{\nu j}$) creates
(annihilates) a photon at the $j$th site on the $\nu $th leg, $\hat{n}_{\nu
j}=\hat{a}_{\nu j}^{\dag }\hat{a}_{\nu j}$ is the number operator, $\omega
_{\nu j}^{0}$ is the qubit frequency, $V_{\nu j}$ is the on-site attractive
interaction at the $j$th site on the $\nu $th leg, $g_{\nu j}$ is the
hopping strength between the nearest-neighbor sites along the leg $\nu $, $%
\tilde{g}_{j}$ is the interleg hopping strength at the rung $j$, and H.c.~is
the Hermitian conjugate. In experiment \cite{YY19}, the transmon qubit has a
strong anharmonicity, $\left\vert V_{\nu j}\right\vert /g_{\nu j}\simeq 20$,
which allows that only one photon can be excited at each site. In such case,
the nonlinear term of the Hamiltonian (\ref{H}) can be safely neglected.

To obtain the wanted synthetic gauge field, here we introduce an ac microwave driving in each transmon qubit, which is experimentally feasible through the flux-bias line \cite{TF}. In this case, each qubit frequency is modulated independently as
\begin{equation}
\omega _{\nu j}(t)=\omega _{\nu j}^{0}+\varepsilon _{\nu j}\sin
\left( u_{\nu j}t+\varphi _{\nu j}\right),  \label{w}
\end{equation}
where $\varepsilon _{\nu j}$, $u_{\nu j}$ and $\varphi _{\nu j}$ are the
driving amplitude, frequency and phase, respectively. By applying the
rotation frame with an unitary operator $\hat{U}=\hat{U}_{1}\times \hat{U}%
_{2}$, where
\begin{subequations}
\begin{eqnarray}
\hat{U}_{1}=\exp \left[-i\sum_{\nu j}\omega _{\nu j}^{0}\hat{n} _{\nu j}t\right],\label{U1}
\end{eqnarray}
\begin{eqnarray}
\hat{U}_{2}=\exp \left[i\sum_{\nu j}\hat{n}_{\nu j}\alpha _{\nu j}\cos
\left( u_{\nu j}t+\varphi _{\nu j}\right) \right], \label{U2}
\end{eqnarray}
\end{subequations}
with $\alpha _{\nu j}=\varepsilon_{\nu j}/u_{\nu j}$, the transformed  Hamiltonian
\begin{eqnarray}\label{trans}
\hat{H}_t=\hat{U}^{\dag }\hat{H}_{BH}\hat{U}+i\frac{d\hat{U}^{\dag }}{dt}\hat{U}
\end{eqnarray}
can be divided into two parts as $\hat{H}_t=\hat{H}_{\text{E}}+\hat{H}_{\text{AB}}$, with
\begin{widetext}
\begin{subequations}
\begin{eqnarray}
\hat{H}_{\text{E}} &=&\underset{\nu }{\sum }g_{\nu 1}\left\{ \hat{a}_{\nu
1}^{\dag }\hat{a}_{\nu 2}e^{-i\Delta _{\nu 2}t}\exp \left[ -i\alpha _{\nu
1}\cos \left( u_{\nu 1}t+\varphi _{\nu 1}\right) \right] \exp \left[ i\alpha
_{\nu 2}\cos \left( u_{\nu 2}t+\varphi _{\nu 2}\right) \right] +\text{H.c.}%
\right\}  \notag \\
&&+\underset{\nu }{\sum }g_{\nu 2}\left\{ \hat{a}_{\nu 2}^{\dag }\hat{a}%
_{\nu 3}e^{-i\Delta _{\nu 3}t}\exp \left[ -i\alpha _{\nu 2}\cos \left(
u_{\nu 2}t+\varphi _{\nu 2}\right) \right] \exp \left[ i\alpha _{\nu 3}\cos
\left( u_{\nu 3}t+\varphi _{\nu 3}\right) \right] +\text{H.c.}\right\} +...,
\end{eqnarray}%
\begin{eqnarray}
\hat{H}_{\text{AB}} &=&\tilde{g}_{1}\left\{ \hat{a}_{A1}^{\dag }\hat{a}%
_{B1}e^{-i\Delta _{AB1}t}\exp \left[ -i\alpha _{A1}\cos \left(
u_{A1}t+\varphi _{A1}\right) \right] \exp \left[ i\alpha _{B1}\cos \left(
u_{B1}t+\varphi _{B1}\right) \right] +\text{H.c.}\right\}  \notag \\
&&+\tilde{g}_{2}\left\{ \hat{a}_{A2}^{\dag }\hat{a}_{B2}e^{-i\Delta
_{AB2}t}\exp \left[ -i\alpha _{A2}\cos \left( u_{A2}t+\varphi _{A2}\right) %
\right] \exp \left[ i\alpha _{B2}\cos \left( u_{B2}t+\varphi _{B2}\right) %
\right] +\text{H.c.}\right\} +...,
\end{eqnarray}%
\end{subequations}
with $\Delta _{\nu j}=\omega _{\nu j}^{0}-\omega _{\nu
j-1}^{0} $ and $\Delta _{ABj}=\omega _{Bj}^{0}-\omega _{Aj}^{0}$.
Using the Jacobi-Anger identity, $\exp \left[ i\alpha \cos \left( ut+\varphi
\right) \right] =\sum_{-\infty }^{\infty }i^{m}J_{m}(\alpha )\exp \left[
im\left( ut+\varphi \right) \right] $, where $J_{m}(\alpha )$ is the Bessel
function of the first kind, we have
\begin{subequations}
\begin{eqnarray}
\hat{H}_{\text{E}} &=&\underset{\nu j}{\sum }g_{\nu j}\hat{a}_{\nu \left(
j-1\right) }^{\dag }\hat{a}_{\nu j}\underset{m_{1}=-\infty }{\overset{\infty
}{\sum }}(-i)^{m_{1}}J_{m_{1}}(\alpha _{\nu \left( j-1\right) })\exp \left[
-im_{1}\left( u_{\nu \left( j-1\right) }t+\varphi _{\nu \left( j-1\right)
}\right) \right]   \notag \\
&&\times \underset{m_{2}=-\infty }{\overset{\infty }{\sum }}%
i^{m_{2}}J_{m_{2}}(\alpha _{\nu j})\exp \left[ i\left( m_{2}u_{\nu j}-\Delta
_{\nu j}\right) t+m_{2}\varphi _{\nu j}\right] +\text{H.c.},  \label{Hvij2}
\end{eqnarray}
\begin{eqnarray}
\hat{H}_{\text{AB}} &=&\underset{j}{\sum }\tilde{g}_{j}\hat{a}_{Aj}^{\dag }%
\hat{a}_{Bj}\underset{m_{1}=-\infty }{\overset{\infty }{\sum }}%
(-i)^{m_{1}}J_{m_{1}}(\alpha _{Aj})\exp \left[ -im_{1}\left( u_{Aj}t+\varphi
_{Aj}\right) \right]   \notag \\
&&\times \underset{m_{2}=-\infty }{\overset{\infty }{\sum }}%
i^{m_{2}}J_{m_{2}}(\alpha _{Bj})\exp \left[ i\left( m_{2}u_{Bj}-\Delta
_{ABj}\right) t+m_{2}\varphi _{Bj}\right] +\text{H.c.}.  \label{HAB2}
\end{eqnarray}%
\end{subequations}
\end{widetext}
When choosing $\Delta _{\nu j}=u_{\nu j}$ ($-u_{\nu j}$) for
odd (even) $j$ and $\Delta _{ABj}=u_{Bj}$, and considering the case that $%
u_{\nu j}\gg \left\{ g_{\nu j},\tilde{g}_{j}\right\} $, the oscillating
terms in Eqs.~(\ref{Hvij2}) and (\ref{HAB2}) are neglected by
applying the rotating-wave approximation. Finally, the effective Hamiltonian
is given by
\begin{eqnarray}
\hat{H} &=&\underset{\nu j}{\sum }\left( t_{\nu j}e^{i\tilde{\varphi}_{\nu
j}}\hat{a}_{\nu \left( j-1\right) }^{\dag }\hat{a}_{\nu j}+\text{H.c.}\right)
\notag \\
&&+\underset{j}{\sum }\left( \tilde{t}_{j}e^{i\tilde{\varphi}_{Bj}}\hat{a}%
_{Aj}^{\dag }\hat{a}_{Bj}+\text{H.c.}\right) ,  \label{H2}
\end{eqnarray}
where $t_{\nu j}=g_{\nu j}J_{0}(\alpha _{\nu j-1})J_{1}(\alpha _{\nu j})$, $%
\tilde{t}_{j}=\tilde{g}_{j}J_{0}(\alpha _{Aj})J_{1}(\alpha _{Bj})$ with $J_{m}(\alpha )$ being the Bessel function of the first kind, and $\tilde{\varphi}_{\nu j}=\left( -1\right) ^{j+1}\varphi _{\nu j}+\pi /2$. 

The Hamiltonian in Eq.~(\ref{H2}) shows clearly that the driving phase $\tilde{\varphi}_{\nu j}$ leads to complex hopping between any two nearest-neighbor sites, and thus make each plaquette accumulate a gauge-invariant magnetic flux $ \varphi _{j}=\tilde{\varphi}_{Aj} -\tilde{\varphi}_{B\left( j-1\right) }\ $, see Fig.~\ref{Two}(b). This synthetic magnetic flux per plaquette can be tuned independently by choosing the driving phases in the transmon qubits, which is better than the previous realizations in other systems. If choosing $\varphi _{j}=\varphi $, the uniform flux is formed \cite{DH14, MP15,ATOM-MA2}; if $\varphi _{j}=(-1)^{j}\varphi $, the staggered flux is generated \cite{RS18,ATOM-MA,LK10,GM10,MA13}; if $\varphi
_{j}=j\varphi $, the site-dependent flux is achieved \cite{ATOM-MA1,PC1,ATOM-BKS,ATOM-HM}.\newline

\section{Ground-state chiral currents}\label{section3}

The synthetic magnetic flux achieved can generate rich quantum phenomena. As
an example, we investigate the experimentally-measurable ground-state chiral
currents and chiral dynamics of the ladder system. For simplicity, we set $%
\alpha _{\nu j}=\alpha $ and $g_{\nu j}=\tilde{g}_{j}=g$, which mean that $%
t_{\nu j}=\tilde{t}_{j}=t_{0}=gJ_{0}(\alpha )J_{1}(\alpha )$. The driving
phases are taken as $\tilde{\varphi}_{Aj}=-\tilde{\varphi}_{Bj}=\varphi /2$,
and the synthetic magnetic flux thus becomes $\varphi _{j}=\varphi $. In
this section, we mainly discuss the ground-state chiral currents. The case
of single-qubit excitation is firstly considered and the case of two-qubit
excitation is then addressed briefly.

By performing the Fourier transformation $\hat{a}_{\nu k}=\sum_{j}e^{ikj}%
\hat{a}_{\nu j}/\sqrt{N}$, where $N$ is the number of the ladder rungs, the
Hamiltonian in Eq.~(\ref{H2}) becomes $\hat{H}=\sum_{k}(\hat{a}_{Ak}^{\dag },\hat{a}%
_{Bk}^{\dag })\hat{h}(k)\left(
\begin{array}{c}
\hat{a}_{Ak} \\
\hat{a}_{Bk}%
\end{array}%
\right) $, where
\begin{equation}
\hat{h}(k)=\varepsilon _{0}(k)\mathit{\hat{I}}+t_{0}\hat{\sigma}%
_{x}+\varepsilon _{z}(k)\hat{\sigma}_{z}.  \label{MTH}
\end{equation}%
where $\mathit{\hat{I}}$ is the identity operator, $\hat{\sigma}_{x}$ and $\hat{%
\sigma}_{z}$ are the Pauli spin operators in the $x$ and $z$ directions, $%
\varepsilon _{0}(k)=2t_{0}\cos (\varphi /2)\cos k$, and $\ \varepsilon
_{z}(k)=2t_{0}\sin (\varphi /2)\sin \left( k\right) $. Since the $A$ and $B$
legs act respectively as the spin-up and spin-down components, the $\hat{%
\sigma}_{x}$ term governs the tunneling between two legs. While the $\hat{%
\sigma}_{z}$ term, determined by the non-zero magnetic flux $\varphi $,
generates spin-momentum locking that the spin-up and spin-down photons
minimize their energies by having the positive and negative momenta,
respectively. The Hamiltonian in Eq.~(\ref{MTH}) exhibits the time-reversal
invariance \cite{DH14}, which leads to the Kramers degeneracy of the ground
state, as will be shown below.

\begin{figure}[t]
\includegraphics[width=8cm]{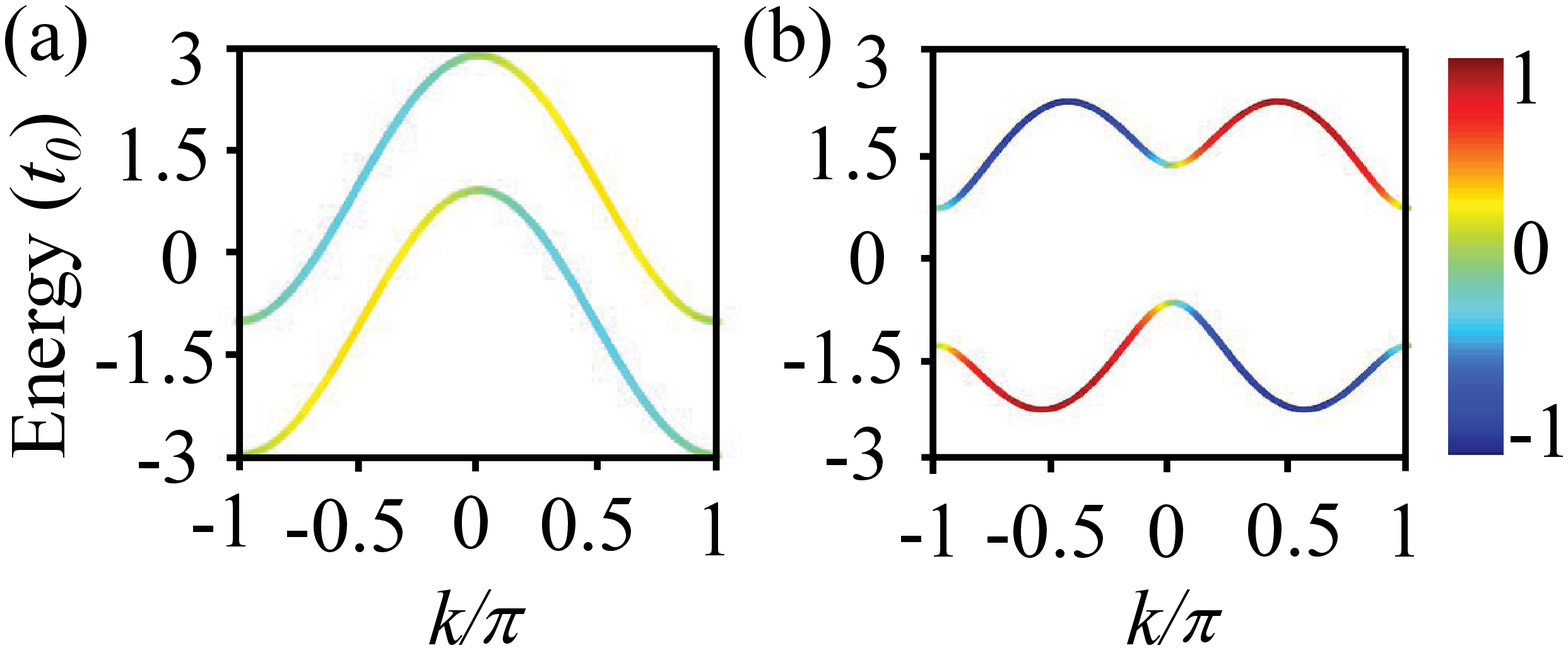}\newline
\caption{The energy bands as functions of the momentum $k$ for (a) $\protect%
\varphi =0.1\protect\pi $ and (b) $\protect\varphi =0.9\protect\pi $. The
color indicates the value of $\left\langle \hat{\protect\sigma}%
_{z}\right\rangle _{k}$. }
\label{Dis}
\end{figure}

With the diagonalization of the momentum-space Hamiltonian in Eq.~(\ref{MTH}), we
obtain two energy bands
\begin{equation}
E(k)=\varepsilon _{0}(k)\pm \sqrt{\varepsilon _{z}(k)^{2}+t_{0}^{2}},
\label{EB}
\end{equation}%
which are plotted, in Fig.~\ref{Dis}, as functions of the momentum $k $ for (a) $\varphi =0.1\pi $ and (b) $\varphi =0.9\pi $. For small $\varphi $, the lower energy band only has one minimum at $k=\pi $ or $k=-\pi$, as shown in Fig.~\ref{Dis}(a). With increasing $\varphi $, two Kramers
degeneracy points occur at $k=\pm q$, with
\begin{equation}
q=\frac{1}{2}\arccos \left\{ \frac{1+\cos \left( \varphi \right) }{2\left[
1-\cos \left( \varphi \right) \right] }+\cos \left( \varphi \right) \right\}
,  \label{km}
\end{equation}%
as shown in Fig.~\ref{Dis}(b). The critical point that the lower energy band
changes from one minimum to two minima is given by $\varphi _{c}=2\arccos (%
\sqrt{17}/4-1/4)$. For simplicity, here we choose $\varphi \in (0,\pi ]$.

With single-qubit excitation, the eigenfunction of the lower energy band is
obtained by%
\begin{equation}
\left\vert \psi _{Lk}\right\rangle =\left( \alpha _{Lk}\hat{a}_{kA}^{\dag
}+\beta _{Lk}\hat{a}_{kB}^{\dag }\right) \left\vert 0\right\rangle ,
\label{GW}
\end{equation}%
where $\left\vert 0\right\rangle $ is the vacuum state,%
\begin{subequations}
\begin{eqnarray}
\alpha _{Lk} &=&\frac{(\bar{\varepsilon}_{z}-\sqrt{1+\bar{\varepsilon}%
_{z}^{2}})}{\sqrt{[(\bar{\varepsilon}_{z}-\sqrt{1+\bar{\varepsilon}_{z}^{2}}%
)^{2}+1]}},  \label{P1}
\end{eqnarray}
\begin{eqnarray}
\beta _{Lk} &=&\frac{1}{\sqrt{[(\bar{\varepsilon}_{z}-\sqrt{1+\bar{%
\varepsilon}_{z}^{2}})^{2}+1]}},  \label{P2}
\end{eqnarray}
\end{subequations}
with $\bar{\varepsilon}_{z}=2\sin (\varphi /2)\sin k$. In terms of Eq.~(\ref%
{GW}), we have
\begin{equation}
\left\langle \hat{\sigma}_{z}\right\rangle _{Lk}=\frac{\left[ \bar{%
\varepsilon}_{z}-\sqrt{1+\bar{\varepsilon}_{z}^{2}}\right] ^{2}-1}{\left[
\bar{\varepsilon}_{z}-\sqrt{1+\bar{\varepsilon}_{z}^{2}}\right] ^{2}+1}.
\label{PP}
\end{equation}%
Equation (\ref{PP}) shows clearly that when $k>0$ (i.e., $\bar{\varepsilon}%
_{z}>0$), $\left\langle \hat{\sigma}_{z}\right\rangle _{Lk}<0$, and vice
versa, which indicates the spin-momentum locking effect induced by the
non-zero magnetic flux. When $\varphi =0$, $\left\langle \hat{\sigma}%
_{z}\right\rangle _{Lk}\equiv 0$ for any $k$.

Due to the spin-momentum locking, the photons in the $A$ leg move towards
the left, whereas the photons in the $B$ leg move towards the right. As a
result, the ladder system with non-zero magnetic flux exhibits a chiral
current defined as
\begin{equation}
\hat{J}_{C}=\hat{J}_{A}-\hat{J}_{B},  \label{CC}
\end{equation}%
with $\hat{J}_{A}=\sum_{j}\hat{J}_{Aj}$ and $\hat{J}_{B}=\sum_{j}\hat{J}%
_{Bj} $, where
\begin{subequations}
\begin{eqnarray}
\hat{J}_{Aj} &=&it_{0}e^{i\frac{\varphi }{2}}\hat{a}_{Aj}^{\dag }\hat{a}%
_{A\left( j+1\right) }+\text{H.c.,}  \label{JA}
\end{eqnarray}
\begin{eqnarray}
\hat{J}_{Bj} &=&it_{0}e^{-i\frac{\varphi }{2}}\hat{a}_{Bj}^{\dag }\hat{a}%
_{B\left( j+1\right) }+\text{H.c..}  \label{JB}
\end{eqnarray}%
\end{subequations}
On the other hand, since the $\hat{\sigma}_{x}$ term in the Hamiltonian in Eq.~(\ref%
{MTH}) governs the tunneling between two legs, it is necessary to define the
current at the rung $j$ as
\begin{equation}
\hat{J}_{j}=it_{0}\hat{a}_{Aj}^{\dag }\hat{a}_{Bj}+\text{H.c.}.  \label{JAB}
\end{equation}

In terms of Eqs.~(\ref{CC})-(\ref{JAB}), we can investigate the ground-state
currents. For $\varphi <\varphi _{c}$, i.e., the lower energy band has one
minimum, the ground-state wavefunction $\left\vert \psi _{\text{G}%
}\right\rangle =\left\vert \psi _{L\pi }\right\rangle $, whose corresponding
currents are calculated as
\begin{subequations}
\begin{eqnarray}
\left\langle \hat{J}_{Aj}\right\rangle _{L\pi } &=&-\left\langle \hat{J}%
_{Bj}\right\rangle _{L\pi }=\frac{t_{0}}{N}\sin (\frac{\varphi }{2}),
\label{JA1}
\end{eqnarray}
\begin{eqnarray}
\left\langle \hat{J}_{C}\right\rangle _{L\pi } &=&2t_{0}\sin (\frac{\varphi
}{2}),  \label{JC1}
\end{eqnarray}
\begin{eqnarray}
\left\langle \hat{J}_{j}\right\rangle _{L\pi } &=&0.  \label{JI1}
\end{eqnarray}
\end{subequations}
These equations show that the currents along the $A$ and $B$ legs have
opposite directions but with the same magnitudes, i.e., a non-zero chiral
current is generated, while the current at the rung $j$ vanishes. This
phenomenon clearly characterizes the Meissner effect \cite{EO01}.

\begin{figure}[t]
\includegraphics[width=8cm]{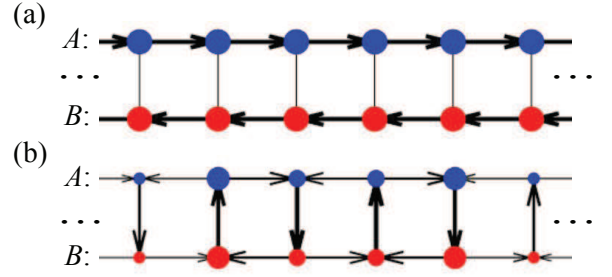}
\caption{The currents between neighboring sites for (a) $%
\protect\varphi =0.1\protect\pi $ and (b) $\protect\varphi =0.9\protect\pi $
when $N=50$. The thicknesses of the arrows indicate their strengths. The
sizes of the blue and red solid spheres denote the local densities at rungs
on the $A$ and $B$ legs, respectively.}
\label{current}
\end{figure}

For $\varphi >\varphi _{c}$, i.e., the lower energy band has two minima, the
ground-state wavefunction becomes $\left\vert \psi _{\text{G}}\right\rangle
=\left( \left\vert \psi _{L\left( -q\right) }\right\rangle +\left\vert \psi
_{Lq}\right\rangle \right) /\sqrt{2}$. In this case, the ground-state
currents are given by
\begin{subequations}
\begin{eqnarray}
\left\langle \hat{J}_{Aj}\right\rangle _{Lq} &=&\frac{t_{0}}{N}\left[ \alpha
_{L(-q)}^{2}\sin \left( q-\frac{\varphi }{2}\right) -\alpha _{Lq}^{2}\sin
\left( q+\frac{\varphi }{2}\right) \right.\notag\\
&& \left. -2\alpha _{L(-q)}\alpha _{Lq}\sin \left(
\frac{\varphi }{2}\right) \cos \left( q+2qj\right) \right] ,  \label{JAQ}
\end{eqnarray}
\begin{eqnarray}
\left\langle \hat{J}_{Bj}\right\rangle _{Lq} &=&\frac{t_{0}}{N}\left[ \beta
_{L(-q)}^{2}\sin \left( q+\frac{\varphi }{2}\right) -\beta _{Lq}^{2}\sin
\left( q-\frac{\varphi }{2}\right)\right. \notag\\
&& \left.+2\beta _{L(-q)}\beta _{Lq}\sin \left(\frac{\varphi }{2}\right)
\cos \left( q+2qj\right) \right] ,  \label{JBQ}
\end{eqnarray}
\begin{eqnarray}
\left\langle \hat{J}_{C}\right\rangle _{Lq} &=&t_{0}\left[ \alpha
_{Lq}^{2}\sin \left( \frac{\varphi }{2}+q\right) -\alpha _{L(-q)}^{2}\sin
\left( q-\frac{\varphi }{2}\right) \right.\notag\\
&&\left. -\beta _{Lq}^{2}\sin \left( q-\frac{\varphi }{2}\right) +\beta _{L(-q)}^{2}\sin \left( \frac{\varphi }{2}+q\right) \right] , \notag\\ \label{JC2}
\end{eqnarray}
\begin{eqnarray}
\left\langle \hat{J}_{j}\right\rangle _{Lq} &=&\frac{t_{0}}{N}\sin \left(
2qj\right) \left[ \alpha _{Lq}\beta _{L(-q)}-\alpha _{L(-q)}\beta _{Lq}%
\right]. \notag\\ \label{JJQ}
\end{eqnarray}%
\end{subequations}
Since $\alpha _{L(-q)}=-\beta _{Lq}$ and $\alpha _{Lq}=-\beta
_{L(-q)}$, it is easy to verify that $\left\langle \hat{J}_{Aj}\right\rangle
_{Lq}=$ $-\left\langle \hat{J}_{Bj}\right\rangle _{Lq}$. Equations (\ref{JAQ}%
), (\ref{JBQ}) and (\ref{JJQ}) show that the currents between any two
nearest-neighbor sites vary periodically when increasing $j$, characterizing
the vortex current \cite{EO17}.

\begin{figure}[t]
\includegraphics[width=8cm]{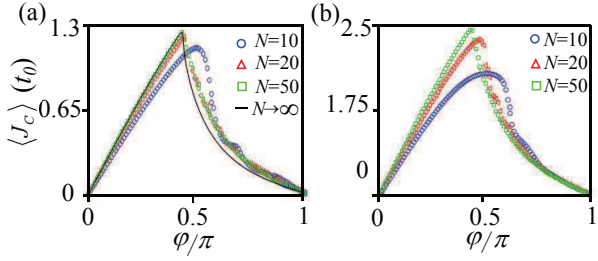}
\caption{The ground-state chiral currents 
as functions of $\protect\varphi $ for (a) single-qubit and (b) two-qubit excitations. The solid line in (a) is the analytical result from Eq.~(\protect\ref{JC2}). The hollow blue circles, red
triangles and green squares are the results of $N=10$, 20 and 50, respectively. }
\label{phase}
\end{figure}

In Fig.~\ref{current}, we plot the currents between any two nearest-neighbor
sites for (a) $\varphi =0.1\pi $\ and (b) $\varphi =0.9\pi $, which support
the above analytical results. While in Fig.~\ref{phase}(a), we plot the
ground-state chiral current as a
function of $\varphi $. With increasing $\varphi $, this chiral current firstly increases to a maximal value at the critical
point $\varphi _{c}$ and then decreases, which characterizes a transition
from the Meissner phase to the vortex phase \cite{MP15,RS18}. For a finite
size, this transition feature still remains but the critical point changes
slightly, which means that the Meissner and vortex phases as well as their
transition can be observed in current experimental setup. In Fig.~\ref{phase}%
(b), we plot the ground-state chiral current for two-qubit excitations,
which has similar properties as those with single-qubit excitation.\newline

\section{Chiral dynamics}\label{section4}

\begin{figure}[t]
\includegraphics[width=8cm]{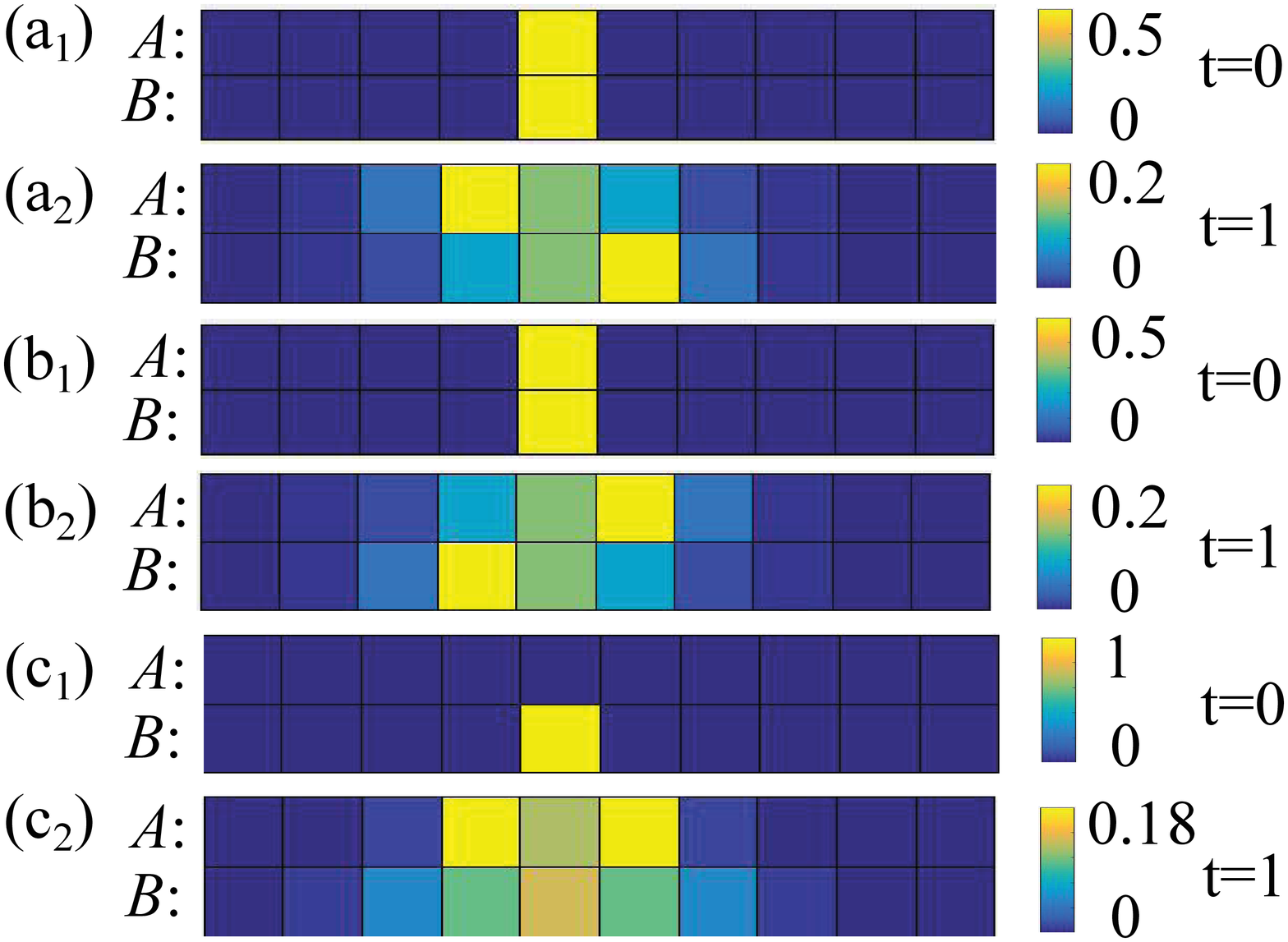}
\caption{The density distributions of photons at each site with single-qubit
excitation for the different initial states $\left\vert \protect\psi %
(0)\right\rangle _{1\text{S}}$ $=(\hat{a}_{A5}^{\dag }+\hat{a}_{B5}^{\dag })$
$\left\vert 0\right\rangle /\protect\sqrt{2}$ with (a$_{1}$) $t=0$ and (a$%
_{2}$) $t=1$, $\left\vert \protect\psi (0)\right\rangle _{1\text{AS}}=(\hat{a%
}_{A5}^{\dag }-\hat{a}_{B5}^{\dag })\left\vert 0\right\rangle /\protect\sqrt{%
2}$ with (b$_{1}$) $t=0$ and (b$_{2}$) $t=1$, and $\left\vert \protect\psi %
(0)\right\rangle _{1\text{E}}=\hat{a}_{B5}^{\dag }\left\vert 0\right\rangle $
with (c$_{1}$) $t=0$ and (c$_{2}$) $t=1$. In all subfigures, the synthetic
magnetic flux $\protect\varphi =0.5\protect\pi $. }
\label{times}
\end{figure}

We now investigate the chiral dynamics of the Hamiltonian in Eq.~(\ref{H2}) with $N=10$ and $\varphi =0.5\pi $. For simplify, we set $t_{0}=1$ in the
following discussion. We first consider the case of single-qubit excitation
denoted by $\hat{a}_{\nu j}^{\dag }\left\vert 0\right\rangle $, which
describes that the qubit at the $j$th site on the$\ \nu $th leg is excited.
In Fig.~\ref{times}, we plot the density distributions of photons at each
site of the ladder for $t=0$ and $t=1$, when the initial states are prepared
respectively as $\left\vert \psi (0)\right\rangle _{1\text{S}}=(\hat{a}%
_{A5}^{\dag }+\hat{a}_{B5}^{\dag })$ $\left\vert 0\right\rangle /\sqrt{2}$ (a%
$_{1}$,a$_{2}$) , $\left\vert \psi (0)\right\rangle _{1\text{AS}}=(\hat{a}%
_{A5}^{\dag }-\hat{a}_{B5}^{\dag })\left\vert 0\right\rangle /\sqrt{2}$ (b$%
_{1}$,b$_{2}$), and $\left\vert \psi (0)\right\rangle _{1\text{E}}=\hat{a}%
_{B5}^{\dag }\left\vert 0\right\rangle $ (c$_{1}$,c$_{2}$). This figure
shows that for the initial state $\left\vert \psi (0)\right\rangle _{1\text{S%
}}$, the most photons move to the left (right) of the central rung on the $A$%
\ $(B)$\ leg [see Fig.~\ref{times}(a$_{2}$)], which characterizes chiral
dynamics. The converse occurs for the initial state $\left\vert \psi
(0)\right\rangle _{1\text{AS}}$ [see Fig.~\ref{times}(b$_{2}$)]. While for
the initial state $\left\vert \psi (0)\right\rangle _{1\text{E}}$, the
photons simultaneously move to the both sides of the central rung [see Fig.~%
\ref{times}(c$_{2}$)], i.e., the chiral dynamics disappears.
In order to see these results clearly, we consider  a short-time ($\delta t$) dynamics, in
which the time-dependent wavefunction is obtained, up to second order, as $%
\left\vert \psi (\delta t)\right\rangle \simeq \lbrack 1-i\hat{H}\delta t-(%
\hat{H}\delta t)^{2}/2]\left\vert \psi (0)\right\rangle $. As a result, the
differences between the photons moving to the right and left of the center
rung on the $A$ and $B$ legs under these three initial states are
given respectively by
\begin{subequations}
\begin{eqnarray}
\left\langle \Delta \hat{n}_{A}\right\rangle _{1\text{S}} = -\left\langle
\Delta \hat{n}_{B}\right\rangle _{1\text{S}}=\delta t^{3}\sin (\varphi ),
\label{1S}
\end{eqnarray}
\begin{eqnarray}
\left\langle \Delta \hat{n}_{A}\right\rangle _{1\text{AS}} = -\left\langle
\Delta \hat{n}_{B}\right\rangle _{1\text{AS}}=-\delta t^{3}\sin (\varphi ),
\label{1AS}
\end{eqnarray}
\begin{eqnarray}
\left\langle \Delta \hat{n}_{A}\right\rangle _{1\text{E}} = \left\langle
\Delta \hat{n}_{B}\right\rangle _{1\text{E}}\text{ }=0,  \label{SS}
\end{eqnarray}%
\end{subequations}
where $\Delta \hat{n}_{A}=\overset{4}{\sum_{j=1}}\hat{n}_{Aj}-\overset{10}{%
\sum_{j=6}}\hat{n}_{Aj}$ and $\Delta \hat{n}_{B}=\overset{4}{\sum_{j=1}}\hat{%
n}_{Bj}-\overset{10}{\sum_{j=6}}\hat{n}_{Bj}$. Equations (\ref{1S}) and (\ref%
{1AS}) show clearly that for the initial states $\left\vert \psi
(0)\right\rangle _{1\text{S}}$ and $\left\vert \psi (0)\right\rangle _{1%
\text{AS}}$, the opposite differences are raised by non-zero $\varphi $, and
the chiral dynamics can thus be formed. While for the initial state $%
\left\vert \psi (0)\right\rangle _{1\text{E}}$, both differences disappears
[see Eq.~(\ref{SS})], i.e., no chiral dynamics occurs.

\begin{figure}[t]
\includegraphics[width=8cm]{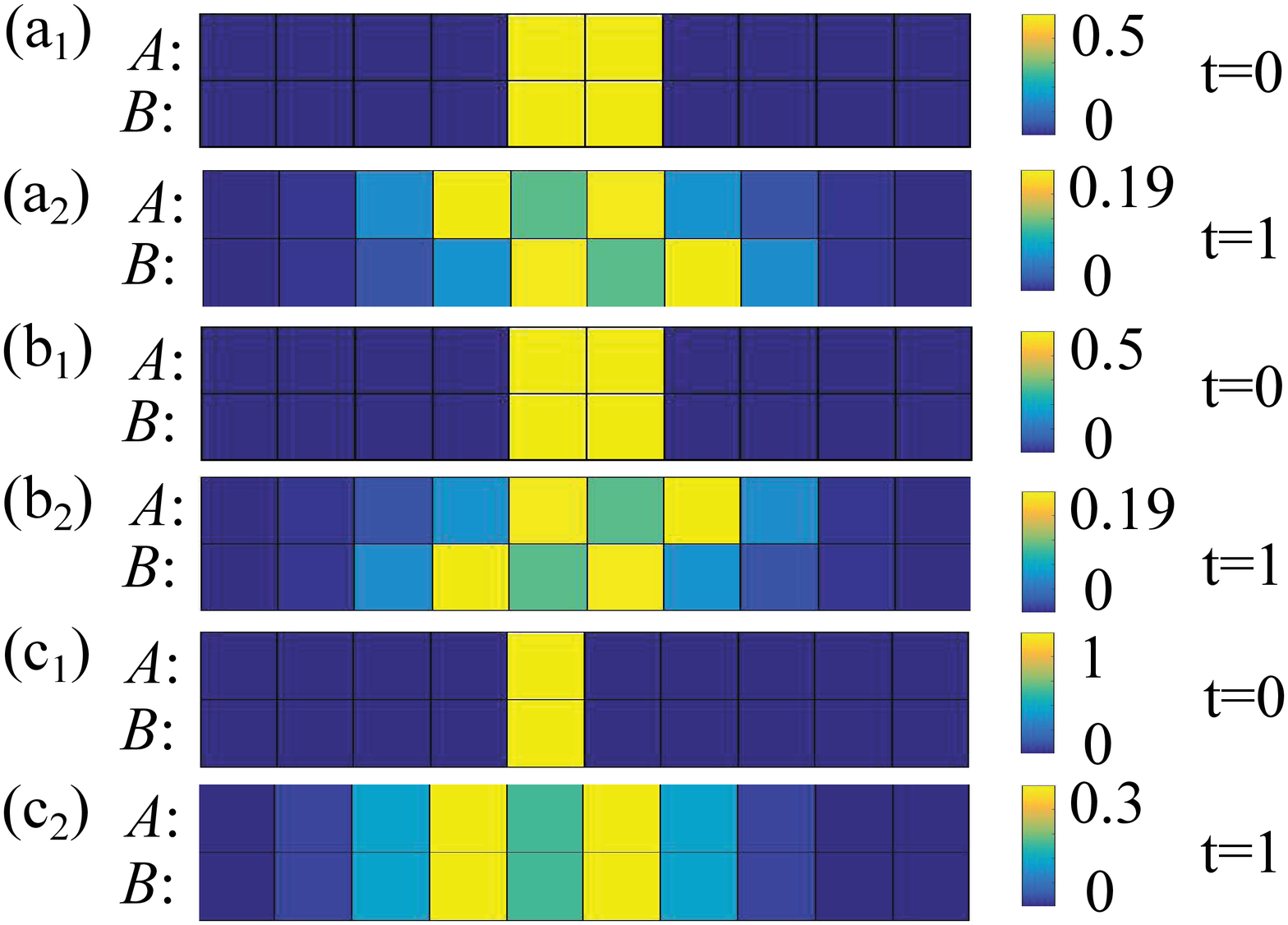}
\caption{The density distributions of photons at each site with two-qubit
excitations for the different initial states $\left\vert \protect\psi %
(0)\right\rangle _{2\text{S}}=(\hat{a}_{A5}^{\dag }+\hat{a}_{B5}^{\dag })(%
\hat{a}_{A6}^{\dag }+\hat{a}_{B6}^{\dag })\left\vert 0\right\rangle /2$ with
(a$_{1}$) $t=0$ and (a$_{2}$) $t=1$, $\left\vert \protect\psi %
(0)\right\rangle _{2\text{AS}}=(\hat{a}_{A5}^{\dag }-\hat{a}_{B5}^{\dag })(%
\hat{a}_{A6}^{\dag }-\hat{a}_{B6}^{\dag })\left\vert 0\right\rangle /2$ with
(b$_{1}$) $t=0$ and (b$_{2}$) $t=1$, and $\left\vert \protect\psi %
(0)\right\rangle _{2\text{E}}=\hat{a}_{A5}^{\dag }\hat{a}_{B5}^{\dag
}\left\vert 0\right\rangle $ with (c$_{1}$) $t=0$ and (c$_{2}$) $t=1$. In
all subfigures, the synthetic magnetic flux $\protect\varphi =0.5\protect\pi
$. }
\label{timet}
\end{figure}

These results can be understood by considering the properties of both two
energy bands shown in Fig.~\ref{Dis}. The fundamental information of the
lower band has been given in the previous section. While for the upper
energy band, its eigenfunction is given by%
\begin{equation}
\left\vert \psi _{Uk}\right\rangle =\left( \alpha _{Uk}\hat{a}_{kA}^{\dag
}+\beta _{Uk}\hat{a}_{kB}^{\dag }\right) \left\vert 0\right\rangle ,
\label{Up}
\end{equation}%
where
\begin{subequations}
\begin{eqnarray}
\alpha _{Uk} &=&\frac{(\bar{\varepsilon}_{z}+\sqrt{1+\bar{\varepsilon}%
_{z}^{2}})}{\sqrt{[(\bar{\varepsilon}_{z}+\sqrt{1+\bar{\varepsilon}_{z}^{2}}%
)^{2}+1]}},
\end{eqnarray}\begin{eqnarray}
\beta _{Uk} &=&\frac{1}{\sqrt{[(\bar{\varepsilon}_{z}+\sqrt{1+\bar{%
\varepsilon}_{z}^{2}})^{2}+1]}}.
\end{eqnarray}
\end{subequations}
From Eqs.~(\ref{GW}) and (\ref{Up}), we obtain%
\begin{subequations}
\begin{eqnarray}
\hat{a}_{Ak}^{\dag }\left\vert 0\right\rangle &=&-\alpha _{L\left( -k\right)
}\left\vert \psi _{Uk}\right\rangle +\alpha _{Lk}\left\vert \psi
_{Lk}\right\rangle ,  \label{Ak}
\end{eqnarray}
\begin{eqnarray}
\hat{a}_{Bk}^{\dag }\left\vert 0\right\rangle &=&-\alpha _{L(-k)}\left\vert
\psi _{Uk}\right\rangle -\alpha _{Lk}\left\vert \psi _{Lk}\right\rangle ,
\label{Bk}
\end{eqnarray}%
\end{subequations}
where the relations $\alpha _{Uk}=\beta _{Lk}$, $\beta _{Uk}=-\alpha _{Lk}$
and $\alpha _{L\left( -k\right) }=-\beta _{Lk}$ have been used. In terms of
Eqs.~(\ref{Ak}) and (\ref{Bk}), the three initial states we have chosen are
rewritten as%
\begin{subequations}
\begin{eqnarray}
\left\vert \psi (0)\right\rangle _{1\text{S}} &=&-\sum_{k}\sqrt{2}%
e^{i5k}\alpha _{L\left( -k\right) }\left\vert \psi _{Uk}\right\rangle ,
\label{S}\end{eqnarray}
\begin{eqnarray}
\left\vert \psi (0)\right\rangle _{1\text{AS}} &=&-\sum_{k}\sqrt{2}%
e^{i5k}\alpha _{Lk}\left\vert \psi _{Lk}\right\rangle ,  \label{AS}
\end{eqnarray}
\begin{eqnarray}
\left\vert \psi (0)\right\rangle _{1\text{E}}\!
&=&\!-\sum_{k}\!e^{i5k}\!\left( \!\alpha _{L(-k)}\!\left\vert \psi
_{Uk}\right\rangle \!+\!\alpha _{Lk}\!\left\vert \psi _{Lk}\right\rangle
\!\right) \!.  \label{SSS}
\end{eqnarray}%
\end{subequations}
Equations (\ref{S}) and (\ref{AS}) show clearly that for the symmetric
(antisymmetric) initial state $\left\vert \psi (0)\right\rangle _{1\text{S}}$
($\left\vert \psi (0)\right\rangle _{1A\text{S}}$), the photons only
populate the upper (lower) band. Due to the spin-momentum locking effect in
the two bands, the chiral dynamics occurs. Equation (\ref{SSS}) shows that
when the initial state is chosen as $\left\vert \psi (0)\right\rangle _{1%
\text{E}}$, the photons populate equally the upper and lower bands with
opposite $k$. Since the two energy bands have opposite chirality (see Fig.~%
\ref{Dis}), the photons move to the both sides of their initial positions
simultaneously and the chiral dynamics thus disappears.

In Fig~\ref{timet}, we plot the density distributions of photons at $t=0$
and $t=1$ for two-qubit excitations of $\hat{a}_{\nu j}^{\dag }\hat{%
a}_{\nu ^{\prime }j^{\prime }}^{\dag }\left\vert 0\right\rangle $, which
describes that the qubits at the $j$th site on the $\nu $th leg and at the $%
j^{\prime }$th site on the $\nu ^{\prime }$th leg are excited. We emphasize
that $\nu $ and $\nu ^{\prime }$, $j$ and $j^{\prime }$ are not equal
simultaneously. The initial states are chosen respectively as $\left\vert
\psi (0)\right\rangle _{2\text{S}}=(\hat{a}_{A5}^{\dag }+\hat{a}_{B5}^{\dag
})(\hat{a}_{A6}^{\dag }+\hat{a}_{B6}^{\dag })\left\vert 0\right\rangle /2$ (a%
$_{1}$,a$_{2}$), $\left\vert \psi (0)\right\rangle _{2\text{AS}}=(\hat{a}%
_{A5}^{\dag }-\hat{a}_{B5}^{\dag })(\hat{a}_{A6}^{\dag }-\hat{a}_{B6}^{\dag
})\left\vert 0\right\rangle /2$ (b$_{1}$,b$_{2}$), and $\left\vert \psi
(0)\right\rangle _{2\text{E}}=\hat{a}_{A5}^{\dag }\hat{a}_{B5}^{\dag
}\left\vert 0\right\rangle $ (c$_{1}$,c$_{2}$). This figure shows the
similar conclusions as those with single-qubit excitation, i.e., for the
initial states $\left\vert \psi (0)\right\rangle _{2\text{S}}$ and $%
\left\vert \psi (0)\right\rangle _{2\text{AS}}$, the system has opposite
chiral dynamics, which disappears for the initial state $\left\vert \psi
(0)\right\rangle _{2\text{E}}$.

\begin{figure}[t]
\includegraphics[width=7cm]{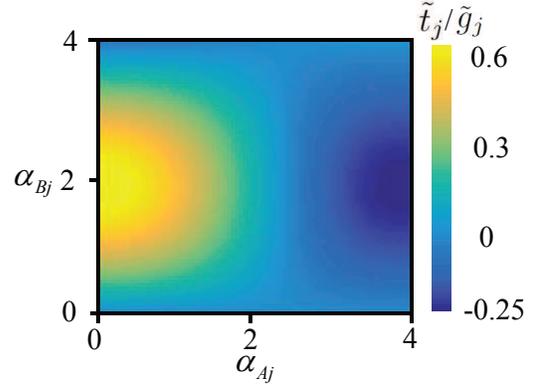}\newline
\caption{The interleg hopping strength as a function of the parameters $%
\protect\alpha _{Aj}$ and $\protect\alpha _{Bj}$.}
\label{Bessel}
\end{figure}

\section{Possible experimental observation}\label{section5}

In this section, we briefly discuss how to detect the ground-state chiral
current and the chiral dynamics in experiments. To observe the ground-state
chiral current, we firstly turn off the parameters $t_{0}$ and $\varphi $
and prepare the corresponding system in its single-qubit (two-qubit)
excitation state $\hat{a}_{\nu j}^{\dag }\left\vert 0\right\rangle $ ($\hat{a%
}_{\nu j}^{\dag }\hat{a}_{\nu ^{\prime }j^{\prime }}^{\dag }\left\vert
0\right\rangle $) by the microwave driving. Then, we adiabatically turn on
these parameters to achieve the ground state of the system by the
Landau-Zener theorem \cite{LZ}. Based on the state tomography that has been
developed successfully \cite{ST}, the density of the states at each site can
be measured and the matrix density $\hat{\rho}$ is thus constructed. This
indicates that the ground-state chiral current can be obtained by
$\left\langle \hat{J}_{C}\right\rangle =tr(\hat{\rho}\hat{J}_{C})$ \cite{PR17}, where $tr$ is the trace operator. If the ground-state chiral current
increases (decreases) as the magnetic flux increases, the Meissner (vortex)
phase is found.

To observe the chiral dynamics, preparing the corresponding initial states
plays a crucial role. The initial state $\hat{a}_{B5}^{\dag }\left\vert
0\right\rangle $ ($\hat{a}_{A5}^{\dag }\hat{a}_{B5}^{\dag }\left\vert
0\right\rangle $) can be prepared directly by the microwave driving. For the
superposition state $(\hat{a}_{A5}^{\dag }+\hat{a}_{B5}^{\dag })$ $%
\left\vert 0\right\rangle /\sqrt{2}$ [$(\hat{a}_{A5}^{\dag }-\hat{a}%
_{B5}^{\dag })\left\vert 0\right\rangle /\sqrt{2}$], we firstly turn off all
the phases and the hopping strengths between the $\nu 5$ site and its
nearest neighbor sites, and prepare the system in the state $\hat{a}_{\nu
5}^{\dag }\left\vert 0\right\rangle $. Since the superposition state $(\hat{a%
}_{A5}^{\dag }+\hat{a}_{B5}^{\dag })$ $\left\vert 0\right\rangle /\sqrt{2}$ [%
$(\hat{a}_{A5}^{\dag }-\hat{a}_{B5}^{\dag })\left\vert 0\right\rangle /\sqrt{%
2}$] is the ground state of the isolate rung subsystem at $j=5$ with the
negative (positive) interleg hopping strength, we adiabatically turn on the
interleg hopping strength towards the negative (positive) by tuning the
parameters $\alpha _{A5}$ and $\alpha _{B5}$ (see Fig.~\ref{Bessel}). As a
result, the required superposition states are prepared. Similarly, in the
case of two-qubit excitations, the initial superposition state $(\hat{a}%
_{A5}^{\dag }+\hat{a}_{B5}^{\dag })(\hat{a}_{A6}^{\dag }+\hat{a}_{B6}^{\dag
})\left\vert 0\right\rangle /2$ [$(\hat{a}_{A5}^{\dag }-\hat{a}_{B5}^{\dag
})(\hat{a}_{A6}^{\dag }-\hat{a}_{B6}^{\dag })\left\vert 0\right\rangle /2$]
is the ground state of the subsystem with two isolate rungs at $j=5$ and $%
j=6 $ with the negative (positive) interleg hopping strengths. We firstly
turn off all the phases and hopping strengths between the $\nu j$ ($j=5$ and
$j=6$) site and its nearest neighbor sites, and drive the system in the
state $\hat{a}_{A5}^{\dag }\hat{a}_{A6}^{\dag }\left\vert 0\right\rangle $.
The required superposition states can be prepared by adiabatically turning
on the interleg hopping strengths both at the $5$th and $6$th rungs via the
parameters $\alpha _{A5}$, $\alpha _{B5}$, $\alpha _{A6}$ and $\alpha _{B6}$
(see Fig.~\ref{Bessel}). Then, the chiral dynamics of the system can be observed by performing quantum state tomography \cite{ST}.\newline

\section{Conclusions}\label{section6}

In summary, we have proposed an experimentally-feasible method to prepare
the synthetic gauge field in the two-leg superconducting circuits with
transmon qubits. In particular, the realized magnetic flux per plaquette is
controlled independently by properly choosing the phases of the
alternating-current microwave driving in each qubit, which is better than
the previous realizations in the other quantum simulation systems. Moreover,
we have obtained the ground-state chiral currents for the single- and
two-qubit excitations and found the Meissner-vortex phase transition. In the
Meissner (vortex) phase, the ground-state chiral current increases
(decreases) as the magnetic flux increases. We have also explored the chiral
dynamics, which depends crucially on the initial state of the system.
Finally, the possible experimental observations of the chiral current and
dynamics are addressed. Our results pave a new route to explore novel
many-body properties \cite{SSN15,AK15,SG16,SG17,RS17,YZ17,RC18,MSC19}, which
can be induced by the interplay of gauge field, two-leg hoppings and
interaction of photons in superconducting circuits.\newline

\acknowledgments

This work is supported partly by the National Key R\&D Program of China
under Grant No.~2017YFA0304203; the NSFC under Grants No.~11674200,
No.~11947226 and No.~11874156; and 1331KSC.

\end{document}